\documentclass{aa}  
 
\usepackage{tabularx}
\usepackage{graphicx}
\usepackage{color}
\usepackage{xcolor}
\usepackage{txfonts}
\usepackage{mathrsfs}
\usepackage{listings}
\usepackage{hyperref}
\usepackage{amsmath}
\usepackage{url}
\usepackage{xspace}
\usepackage{soul}  
\usepackage{placeins}

\usepackage{glossaries}
\setacronymstyle{long-short}
\glsdisablehyper

\newcommand{\muHz}{\,\mu\mathrm{Hz}}
\newcommand{\ppmmuHz}{\,\mathrm{ppm}^2/\mu\mathrm{Hz}}
\newcommand{\K}{\,\mathrm{K}}
\newcommand{\Msun}{\,\mathrm{M}_\odot}
\newcommand{\Rsun}{\,\mathrm{R}_\odot}
\newcommand{\Lsun}{\,\mathrm{L}_\odot}
\newcommand{\Teffsun}{\,\mathrm{T}_{\mathrm{eff},\odot}}
\newcommand{\numaxsun}{\,\mathrm{\nu}_{\mathrm{max},\odot}}

\newcommand{\numax}{\nu_{\mathrm{max}}\xspace}
\newcommand{\dnu}{\Delta\nu\xspace}
\newcommand{\gammaenv}{\Gamma_{\mathrm{E}}}
\newcommand{\gammaenvi}{\Gamma_{\mathrm{E},i}}

\newcommand{\Henv}{H_{\mathrm{E}}}
\newcommand{\Amax}{A_{\mathrm{max}}}
\newcommand{\snr}{S/N\xspace}
\newcommand{\dof}{N_{\mathrm{dof}}\xspace}

\newcommand{\Teff}{\,T_\mathrm{eff}}
\newcommand{\vmag}{\,V_{\mathrm{mag}}}

\newcommand{\pmodes}{p-modes\xspace}
\newcommand{\pmode}{p-mode\xspace}
\newcommand{\epsret}{$\epsilon$~Reticuli\xspace}

\newcommand{\kepler}{\textit{Kepler}\xspace}

\newcommand{\gaia}{Gaia\xspace}

\newcommand{\Nenv}{\,N_{\mathrm{E}}}
\newcommand{\Nenvi}{\,N_{\mathrm{E},i}}

\newacronym{ATL}{ATL}{Asteroseismic Target List}
\newcommand{\ATL}{\gls{ATL}\xspace}

\newacronym{RG}{RG}{red giant}
\newcommand{\RG}{\gls{RG}\xspace}
\newcommand{\RGs}{\glspl{RG}\xspace}

\newacronym{MS}{MS}{main-sequence}
\newcommand{\MS}{\gls{MS}\xspace}

\newacronym{SG}{SG}{subgiant}
\newcommand{\SG}{\gls{SG}\xspace}

\newacronym{FFI}{FFI}{full-frame image}
\newcommand{\FFI}{\gls{FFI}\xspace}

\newacronym{PSD}{PSD}{power spectral density}
\newcommand{\PSD}{\gls{PSD}\xspace}

\newacronym{LC}{LC}{long cadence}
\newcommand{\LC}{\gls{LC}\xspace}

\newacronym{SC}{SC}{short cadence}
\newcommand{\SC}{\gls{SC}\xspace}

\newacronym{ACF}{ACF}{autocorrelation function}
\newcommand{\ACF}{\gls{ACF}\xspace}

\newacronym{TESS}{TESS}{Transit Exoplanet Survey Satellite}
\newcommand{\tess}{\gls{TESS}\xspace}

\newcommand{\corot}{CoRoT\xspace}

\newcommand{\plato}{PLATO\xspace}

\newacronym{SPOC}{SPOC}{Science Processing Operations Center}
\newcommand{\SPOC}{\gls{SPOC}\xspace}

\newacronym{TIC}{TIC}{TESS Input Catalog}
\newcommand{\TIC}{\gls{TIC}\xspace}

\newacronym{HR}{HR}{Hertzsprung-Russell}
\newcommand{\HR}{\gls{HR}\xspace}

\newacronym{FP}{FP}{false positive}
\newcommand{\FP}{\gls{FP}\xspace}

\newacronym{TP}{TP}{true positive}
\newcommand{\TP}{\gls{TP}\xspace}

\newacronym{PE}{PE}{power excess}
\newcommand{\PE}{\gls{PE}\xspace}

\newacronym{RP}{RP}{repeating pattern}
\newcommand{\RP}{\gls{RP}\xspace}

\begin{document}

   \title{A probabilistic method for detecting solar-like oscillations using meaningful prior information}
   \subtitle{Application to TESS 2-minute photometry}

   \author{M. B. Nielsen\inst{1,2,3}
          \and
          E. Hatt\inst{1} 
          \and
          W. J. Chaplin\inst{1,2}
          \and
          W. H. Ball\inst{1,2}
          \and
          G. R. Davies\inst{1}
          }

   \institute{School of Physics and Astronomy, University of Birmingham, Birmingham B15 2TT, UK\\
              \email{m.b.nielsen.1@bham.ac.uk}
              \and
              Stellar Astrophysics Centre (SAC), Department of Physics and Astronomy, Aarhus University, Ny
Munkegade 120, DK-8000 Aarhus C, Denmark
             \and
             Center for Space Science, NYUAD Institute, New York University Abu Dhabi, PO Box 129188, Abu
Dhabi, United Arab Emirates
              }

   \date{Accepted March 9, 2022}

  \abstract
   {Current and future space-based observatories such as the Transiting Exoplanet Survey Satellite (TESS) and PLATO are set to provide an enormous amount of new data on oscillating stars, and in particular stars that oscillate similar to the Sun. Solar-like oscillators constitute the majority of known oscillating stars and so automated analysis methods are becoming an ever increasing necessity to make as much use of these data as possible.}
   {Here we aim to construct an algorithm that can automatically determine if a given time series of photometric measurements shows evidence of solar-like oscillations. The algorithm is aimed at analyzing data from the TESS mission and the future PLATO mission, and in particular stars in the main-sequence and subgiant evolutionary stages.}
   {The algorithm first tests the range of observable frequencies in the power spectrum of a TESS light curve for an excess that is consistent with that expected from solar-like oscillations. In addition, the algorithm tests if a repeating pattern of oscillation frequencies is present in the time series, and whether it is consistent with the large separation seen in solar-like oscillators. Both methods use scaling relations and observations which were established and obtained during the CoRoT, \textit{Kepler}, and K2 missions.}
   {Using a set of test data consisting of visually confirmed solar-like oscillators and nonoscillators observed by TESS, we find that the proposed algorithm can attain a $94.7\%$ \TP rate and a $8.2\%$ \FP rate at peak accuracy. However, by applying stricter selection criteria, the \FP rate can be reduced to $\approx2\%$, while retaining an $80\%$ \TP rate.}
   {}


   \maketitle

\section{Introduction}
The recent increase in availability of high quality data from space-based observatories, such as \emph{MOST} \citep{Walker2003}, \corot \citep{Baglin2009}, and \kepler \citep{Borucki2010}, has allowed for the broad application of asteroseismology in characterizing stellar systems \citep[e.g.,][]{jcd2010, Huber2013, Hall2021} and populations \citep[e.g.,][]{Handberg2017, Montalban2021, Lyttle2021}. The Transiting Exoplanet Survey Satellite \citep[TESS,][]{Ricker2014} was launched with the purpose of detecting close-in planets around bright stars, but as was the case with its most direct predecessor \kepler, the observations from \glsunset{TESS}\tess also reveal a rich variety of oscillators \citep[][]{Antoci2019, Pedersen2019, Campante2019, Metcalfe2020}. 

The \tess mission has so far produced millions of photometric light curves due to the chosen observing strategy \citep[see][]{Huang2020, Kunimoto2021}. \tess has now observed almost the entire sky with a step-and-stare method, producing time series of approximately 27 days at each step, also known as a sector. The orientation of the sector pattern during the nominal mission means that targets closer to the ecliptic poles are typically observed in several sectors and, near the equator, the fields are observed in only one sector.

Several authors have already shown that it is possible to detect solar-like oscillations using \tess data.
\citet{Mackereth2021} found $6388$ oscillating \RGs observed for one year in the southern ecliptic hemisphere, using an early custom reduction of \tess observations \citep{Nardiello2021}. \citet{Stello2022} detected oscillations in approximately $4500$ red giants observed by \tess in the \kepler field, and recently \citet{Hon2021} detected oscillations in a further $158,505$ \RG stars across the fields observed by \tess. So far, however, only a small number of \SG stars \citep[][]{Huber2019, Chaplin2020, Ball2020, Addison2021} and \MS stars \citep{Nielsen2020, Chontos2021} have shown clear detections, owing to them being fainter and to the lower amplitude of the oscillations in these stars compared to \RG stars. Prior to the launch of the mission, \citet{Schofield2019} compiled the \tess \ATL, suggesting that $25,000$ \MS and \SG targets would have greater than a $5\%$ probability of showing solar-like oscillations. However, with \tess currently in its first extended mission and, producing even more light curves, a consistent search for solar-like oscillators requires an automated approach, even when prioritizing the short-listed \ATL targets.

Methods already exist that are well suited for automatically classifying different types of variability in stars \citep[e.g.,][]{Debosscher2011, Armstrong2016, Jamal2020}. Others place more emphasis on detections of specific types of variability, such as oscillations in \RG stars \citep[e.g.,][]{Stello2017, Hon2019, Kuszlewicz2020}. Recently \citet{Audenaert2021} presented an algorithm for classifying several types of variability in specifically single sectors of \tess data including solar-like oscillations. Here we present a new method for evaluating the probability that a light curve consisting of one or more sectors exhibits solar-like oscillations. The signal-to-noise ratio (\snr) of solar-like oscillations increases largely by simply extending the observing time. A method that utilizes all the available sectors of \tess data is therefore necessary to maximize the yield of detected oscillators observed by current missions, but also from future large surveys like the \plato mission \citep[][]{Rauer2014}. 

The methodology presented here focuses on \SG and \MS stars, and we restrict the testing of the method and discussion to the 2 minute cadence \tess data. The sampling rate of the 30 minute cadence of the nominal mission \FFI data is not suitable for detecting the high-frequency oscillations in \SG and \MS stars. The algorithm does not, however, require a particular observation cadence, and so is in principle applicable to \LC time series of \RG stars observed by \tess or other missions. 

The detection algorithm is divided into two parts: first, we estimate the probability that an excess in the power spectrum of a time series is consistent with what is expected from a solar-like oscillator, and is simultaneously inconsistent with the background noise level; secondly, we estimate the probability that the power excess exhibits a repeating pattern of peaks which is consistent with that expected from the frequency separation of modes in solar-like oscillators. 

This paper will focus on the methodology and performance of the algorithm, and the detections and measurements will be presented and cataloged separately in Hatt et al. (2022, in prep). The paper is structured as follows. Section \ref{sec:tgtselection} describes the types of stars that the detection algorithm is focused on, as well as the choice of testing and training sets that are used to calibrate and evaluate the method. Section \ref{sec:data} discusses the choice of data and the initial processing steps that are performed before applying the detection method. In Sect. \ref{sec:method} we describe the two parts of the algorithm, and their individual and combined performance metrics are presented in Sect. \ref{sec:performance}. Finally we discuss some potential improvements and future applications in Sect. \ref{sec:conclusions}.

\section{Target selection} \label{sec:tgtselection}
We define solar-like oscillators as showing only oscillations that are similar to the type observed in the Sun. These oscillations are stochastically excited and damped by convection in the outer stellar envelope, and propagate throughout the stellar interior. The restoring force for these oscillations is the gradient of pressure, and so they are often referred to as \pmodes. 

Since solar-like oscillations are driven by convection, they can appear in any star with a significant convective region, and are not only restricted to solar analogs. Solar-like oscillators are ubiquitous across the cooler part of the \HR diagram in stars with a substantial convection zone, and the main limiting factors for detecting their variability are the target brightness and the duration of the observations. \citet{Yu2018} detected oscillations in approximately $16000$ \RG stars during the nominal \kepler mission, down to a $V$-band magnitude of approximately $15$. On the main sequence solar-like oscillations have been observed in a temperature range of approximately $\Teff=4800-7000\K$. The amplitude of solar-like oscillations increases with the effective surface temperature $\Teff$ and decreases with increasing surface gravity \citep[e.g.,][]{Kjeldsen1995}. Combined with the stars becoming intrinsically dimmer, the detection of solar-like oscillations further down the main sequence therefore becomes more challenging. The faintest detections of solar-like oscillations among \MS stars during the \kepler mission is approximately $\vmag=12$ \citep{Huber2013}. 
 
\begin{figure}
    \centering
    \includegraphics[width=1\columnwidth]{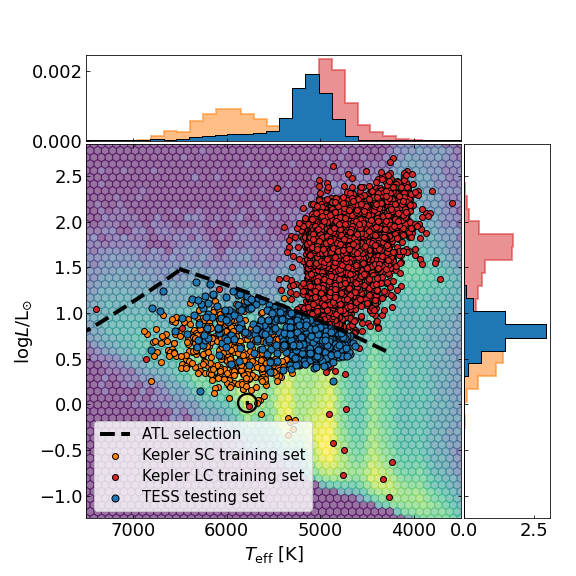}
    \caption{Luminosity and effective temperatures of the \tess testing sets (blue), the \kepler short cadence set (orange), and the \kepler \RG long-cadence set in red. The top and right histograms show the collapsed distributions along each axis. The background hexgrid histogram shows a selection of \gaia stars for reference. }
    \label{fig:hrd}
\end{figure}

 Based on the selection criteria that \citet[][]{Schofield2019} used to construct the \ATL, and the previously observed \kepler stars, we define an approximate range of targets for which our method is optimized. The upper limit of this range is set by $L/\Lsun = \left(\Teff/8907K\right)^{-10.75}$ \citep{Chaplin2011a}, which approximately delineates the $\delta$ Scuti instability strip and the cooler solar-like oscillators \citep{Houdek1999}. For stars cooler than approximately $6500$K, the upper limit in luminosity is instead set by the Nyquist frequency of the \kepler \LC and \tess \FFI images at $\approx282\muHz$. The characteristic time scale of solar-like oscillations can be expressed in terms of the stellar luminosity and effective temperature (see Sect. \ref{sec:PE}), and so the Nyquist frequency limit can be represented as an upper limit in luminosity given approximately  by $L/\Lsun = 16.7 \left(\Teff/\Teffsun\right)^5$ \citep[see][]{Schofield2019}. The resulting cuts in the luminosity and temperature are shown in Fig. \ref{fig:hrd}. 
 
To calibrate the detection method we construct a training set from the \SG and \MS stars observed by \kepler in \SC and a subset of the \RG stars observed in \LC mode. The \SC set is compiled from \citet{White2011}, \citet{SilvaAguirre2015}, \citet{Serenelli2017}, and \citet{Lund2017}, and consists of $495$ targets. We supplement the \SC set by $68$ low-luminosity \RG stars with $\numax>240\muHz$ from the \citet{Yu2018} sample. This bridges the gap between the \kepler \SC sample which is predominantly \MS and \SG stars, and the \LC sample consisting of \RG stars (see Fig. \ref{fig:hrd}). We do not use any of the light curve information from the training set, and only use the global stellar and asteroseismic parameters to calibrate the detection method.  

The intent is to apply the detection algorithm to \tess data, and so as a testing set we use the light curves from 400 solar-like oscillators that were manually identified in the \ATL and that were observed by \tess (see Hatt et al. 2022, in prep). In addition, we compiled a list of 400 targets which we manually verified as nondetections. This sample was drawn from the remainder of the \ATL weighted by a kernel density estimate of the effective temperature distribution from the oscillating sample. This produced a sample of nonoscillating stars with approximately the same distribution in $\Teff$. Since the \ATL targets form a narrow range in luminosity the two testing sets are nearly identically distributed in the \HR diagram. These stars are shown in Fig. \ref{fig:hrd}, and fall in between the two main parts of the \kepler calibration set. 

\begin{figure}
    \centering
    \includegraphics[width=1\columnwidth]{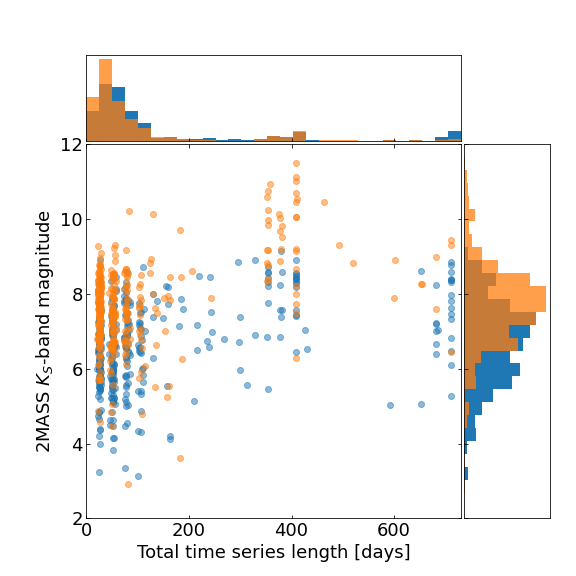}
    \caption{Distribution of 2MASS $K_S$-band magnitudes and time series lengths for the \tess testing set shown in Fig. \ref{fig:hrd}. The testing set consists of a manually validated set of 400 oscillating and 400 nonoscillating stars.}
    \label{fig:testset}
\end{figure}

With the luminosity and effective temperature distributions being approximately equal for the two sets, the remaining governing parameters for the visibility of solar-like oscillations are the apparent brightness and total duration of the observations. Figure \ref{fig:testset} shows the distribution of the oscillating and nonoscillating test sets in $K_S$-band magnitude and total time series length. The nonoscillating set is fainter by approximately one magnitude, and the time series are on average shorter by 36 days, equating to slightly more than one sector. 

The detection methods presented here rely in part on establishing prior constraints on the stellar radius, which in turn can help constrain the characteristic oscillation frequency of the star (see Sect. \ref{sec:PEpriors}). To compute these we use \gaia data release 2 parallaxes, 2MASS $K_S$-band magnitudes, and effective temperatures from version 8 of the TESS Input Catalog \citep[TIC, see][]{Stassun2019}, and bolometric corrections from \citet[][]{Chiavassa2018}. 

\section{Data selection and preparation}
\label{sec:data}

We use the Presearch Data Conditioning (PDC) light curves from \tess data release 42, prepared by the Science Processing Operations Center \citep[SPOC,][]{Jenkins2016}. We assume that the majority of any large-scale instrumental variability has been removed, and the remaining variability is largely intrinsic to the stars.  We use all the available data for each star, concatenating the individual sectors into a single light curve. A small subset of the test set also has 20 second cadence data available, but this is not currently included and we only use the 2-minute data. 

The method makes extensive use of the \PSD of the time series. To compute the \PSD we use the fast Lomb-Scargle method \citep{Lomb1976, Scargle1982} from the \texttt{Astropy} library \citep{astropy2018}. In the following we assume that the power spectrum is critically sampled with a frequency resolution of $\Delta\nu_{\mathrm{T}}$ equivalent to the inverse time series length, and that the frequency bins are independent. Furthermore, the methods rely on the assumption that the noise in each bin is distributed according to a $\chi^2$ distribution with two degrees of freedom ($\dof$). 

However, the \glsunset{SPOC}\SPOC time series can exhibit gaps which cause correlations between frequency bins, and so these approximations are not strictly valid. Gaps appear where either no data are recorded for a target, or a degradation of the data quality has occurred and they are therefore omitted. These events may be caused by, for example, pixel saturation by stray light from solar-system bodies or stochastic cosmic ray hits \citep[see][for details]{Twicken2020}. Periodic gaps are typically caused by data downloads during each orbit. Since \tess has now performed more than two sets of one year observing campaigns for both the southern and northern ecliptic hemispheres, some light curves may have a single large gap spanning up to approximately two years.  

We find that leaving the gaps untreated in the time series produces a significant amount of false positive detections (see Sect. \ref{sec:performance} and Fig. \ref{fig:keplermerit}). We therefore make the rather crude assumption that the variability in the time series from a mode is uncorrelated on either side of a large gap, due to the finite lifetime of the oscillation. The gaps longer than the typical mode lifetime of a given star may then be removed by adjusting the observation time stamps, thereby reducing the correlation between frequency bins in the power spectrum. The characteristic lifetime of solar-like oscillations is on the order of several days to weeks \citep{Corsaro2012, Appourchaux2014, Lund2017}, and decreases as a function of effective temperature and the characteristic mode frequencies in a star. With the testing sets used here, we find that adjusting the time stamps of the observations to remove gaps longer than $50$ day reduces the \snr of the modes in the \PSD, but strongly reduces the number of false positive detections. The large gaps are reduced to the two minute cadence, while shorter gaps are unaltered. 

\section{Detection methods}
\label{sec:method}
\begin{figure*}
    \centering
    \includegraphics[width=1\textwidth]{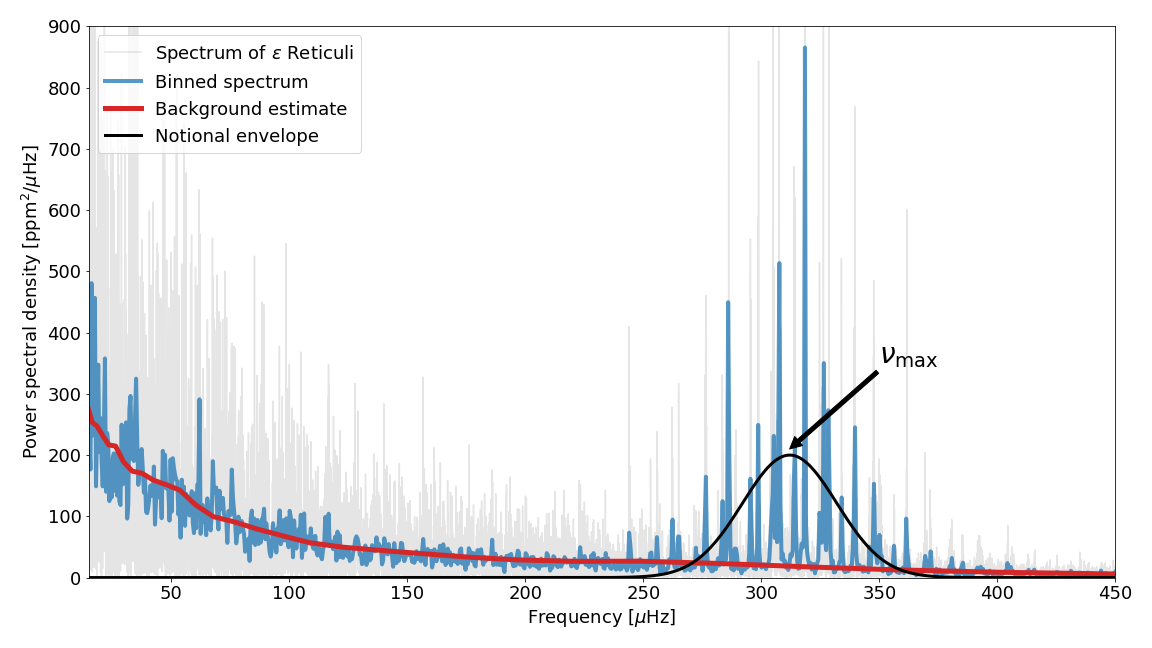}
    \caption{Example power spectrum of \epsret (TIC198079199) shown in gray, with the spectrum binned to $0.2\muHz$ in blue. The estimate of the correlated and white noise background is shown in red. The power in the \pmode envelope is distributed approximately according to a Gaussian (black) centered on $\numax$.}
    \label{fig:example_psd}
\end{figure*}

The detection methods described here rely on identifying the two main characteristics of the variability in a solar-like oscillator. The first is that the power of the oscillation modes in a star appears roughly according to a Gaussian centered on the characteristic oscillation frequency $\numax$. Figure \ref{fig:example_psd} shows the spectrum of \epsret, where the power in the oscillation modes gradually tapers off for modes far from $\numax$.

The individual oscillation modes of a star are characterized in terms of spherical harmonics with angular degree $l$ and azimuthal order $m$, along with a radial order $n$ identifying the overtone number of a particular mode. The second main characteristic of solar-like oscillations is that modes of the same angular degree, but consecutive radial order, are approximately equally spaced in frequency by the large separation $\dnu$. This gives rise to the regular pattern of the modes in the \pmode envelope. 
 
Both of these envelope characteristics scale to a large extent according to just $\numax$ and $\Teff$ \citep[see, e.g., ][]{Chaplin2011a}, giving rise to a set of approximate scaling relations for power in the variability of solar-like oscillators. Given a value of $\Teff$ for a star we can therefore establish a "notional" \pmode envelope with a $\numax$ equivalent to any test frequency bin in the spectrum, and with a height, width, and large separation determined by the scaling relations. We use this in the following to compare the observed variability to what is expected from a solar-like oscillator.

The first method we apply uses scaling relations for the envelope power to evaluate the probability that a \PE is due to solar-like oscillations. The second method uses the autocorrelation of the time series to search for correlated variability that is consistent with the \RP of the large frequency separation in solar-like oscillators. 

\subsection{The power excess detection method}
\label{sec:PE}
The \PE method evaluates the probabilities of two hypotheses: first, the $\mathrm{H}_0$ hypothesis asks if the power in a range around a test frequency bin is consistent with the background noise level in the spectrum; second, the $\mathrm{H}_1$ hypothesis asks if the power in that same range is consistent with solar-like oscillations. Using the probability of the $\mathrm{H}_0$ hypothesis we can identify a power excess in the spectrum, that is, frequency bins where the probability of $\mathrm{H}_0$ is low. With the $\mathrm{H}_1$ hypothesis we evaluate the probability that this excess is equivalent to what we expect an actual \pmode envelope to look like, and not another source of variability in the time series. These could be other types of oscillators, exoplanet transits, rotational variability, or uncorrected systematic effects. The combination of these two probabilities is used to evaluate the posterior probability that a \pmode envelope is present and centered at a particular test frequency.

\subsubsection{Evaluating the $\mathrm{H}_0$ probability} 
Solar-like oscillations typically appear as several consecutive overtones in the high \snr cases (see Fig. \ref{fig:example_psd}), or a broad range of excess power in the low \snr cases. Several consecutive bins of low \snr variability may therefore still indicate the presence of an envelope. So  rather than simply evaluating the probability of observing an excess at a single frequency, it is more prudent to evaluate the joint probability over a range of frequencies surrounding the test frequency. 

A \pmode envelope is expected to appear as a roughly Gaussian distribution of power, centered on a frequency $\numax$, with a full width at half maximum of
\begin{equation}
    \gammaenv =\begin{cases}0.66\left(\frac{\numax}{\mu\mathrm{Hz}}\right)^{0.88} & \Teff \leq 5600K\\ 
                    0.66\left(\frac{\numax}{\mu\mathrm{Hz}}\right)^{0.88}(1+6\times10^{-4}(\Teff-\Teffsun)) & \Teff > 5600K
        \end{cases},
    \label{eq:gammaenv}
\end{equation}
and we therefore sum the power within a range of $\numax\pm\gammaenv/2$. The parameters of Eq. \ref{eq:gammaenv} are derived from fits to the power spectra of \kepler targets by \citet{Mosser2012b}, with an additional temperature dependence appearing for hotter stars as determined by \citet{SchofieldThesis}.

Since we are testing the noise characteristics in a range around the test frequency by summing the power, the likelihood of $\mathrm{H}_0$ is given by a $\chi^2$ distribution with $\dof=2\Nenv$ degrees of freedom. Here, $\Nenv$ is the number of frequency bins in the notional \pmode envelope, and is given by $\Nenv=\gammaenv/\Delta\nu_{T}$. For a given frequency bin we can then compute the likelihood of $\mathrm{H}_0$ by
\begin{equation}
\mathcal{L}(S_i|\mathrm{H}_0) = \chi^2_{2\Nenvi}(S_i), 
\label{eq:H0prob}
\end{equation}
where 
\begin{equation}
\chi^2_{2\Nenvi}(S_i) = \frac{S_i^{2\Nenvi-1}}{\gamma(2\Nenvi)}\exp(-S_i), 
\end{equation}
and $S_i=\sum_{i-\Nenvi/2}^{i+\Nenvi/2}p_i/b_i$ is the sum of the \snr given by the \PSD, $p$, and the background noise estimate $b$. The sum is over the range of the spectrum that encompasses the full-width at half maximum of the notional envelope. 
 
The background noise level, $b$, for a typical solar-like oscillator consists of a frequency-independent noise term (called white noise or shot noise) and a number of frequency-dependent red noise terms caused by granulation and other long-period brightness fluctuations on the stellar surface. These background terms can be modeled in the \PSD using a sum of zero-frequency Lorentzian profiles \citep{Harvey1985} or more flexible Lorentzian-like models \citep{Michel2009, Karoff2012}. While such an approach accounts for much of the background noise in the spectrum, and can yield information about the physical properties of the star \citep{Bastien2016, Bugnet2018, Kallinger2019}, the choice of the model and number of background terms is not always clear \citep[see, e.g.,][]{Kallinger2014}.

We therefore use a nonparametric method which approximates the background by computing the median power around a series of separated frequency bins in the spectrum. These bins are evenly spaced in log-frequency from the lowest observed frequency in the spectrum to the Nyquist frequency. We let the range around each bin follow the width of a notional \pmode envelope at that frequency  (Eq. \ref{eq:gammaenv}). A linear interpolation is then used to represent the background variation between these points. Figure \ref{fig:example_psd} shows the background estimate for an example star \epsret. This estimate follows the granulation terms on short frequency scales at low frequencies, while becoming almost constant at large frequencies where the background noise is almost entirely due to the shot noise. 

\subsubsection{Evaluating the $\mathrm{H}_1$ probability} 
The $\mathrm{H}_0$ hypothesis is useful for detecting a power excess in the spectrum. However, we are also interested in how well this excess compares to what we expect a \pmode envelope to look like. At each test frequency we therefore compare the total power in a range given by $\gammaenv$, as above, with the sum of the power predicted by asteroseismic scaling relations. 

Similar to the envelope width, a scaling relation can be derived for the predicted power $p_{\mathrm{pred}}$ of the envelope, which we use to compare to the observed power. The details of computing $p_{\mathrm{pred}}$ follows that of 
 \citet{Chaplin2011a} and \citet{Schofield2019}, and we present the derivation in Appendix \ref{app:ppred} \citep[see also][chapter 5]{Basu2017}. 

Given the observed \PSD, $p$, and the predicted \PSD, $p_{\mathrm{pred}}$, the likelihood of the $\mathrm{H}_1$ hypothesis can then be computed by
 
\begin{equation}
    \mathcal{L}(R_i|\mathrm{H}_1) = \chi^2_{2\Nenvi} (R_i),
    \label{eq:H1prob}
\end{equation}
where $R_i=\sum_{i-\Nenvi/2}^{i+\Nenvi/2}p_i/p_{\mathrm{pred},i}$. 

Using Eq. \ref{eq:H0prob} and \ref{eq:H1prob} allows us to compute the posterior probability of the $\mathrm{H}_1$ hypothesis
\begin{equation}
P_{\mathrm{PE},i} =
\int_{A}
\frac{P_{\mathrm{H}_1,i}}{P_{\mathrm{H}_1,i} + P_{\mathrm{H}_0,i}}dA^{\prime},
\label{eq:PEposterior}
\end{equation}
where
\begin{equation}
P_{\mathrm{H}_1,i} = \mathcal{P}_{\mathrm{H}_1,i}\mathcal{L}(R_i|\mathrm{H}_1),    
\end{equation}
and
\begin{equation}
P_{\mathrm{H}_0,i} =   \mathcal{P}_{\mathrm{H_0},i}\mathcal{L}(S_i|H_0).
\end{equation}

Each term has an associated prior probability, $\mathcal{P}(\mathrm{H}_0)$ and $\mathcal{P}(\mathrm{H}_1)$, and the integral over $A^{\prime}$ (see Appendix \ref{app:ppred}) marginalizes over the uncertainty in the envelope amplitude scaling relation. 

\begin{figure*}
    \centering
    \includegraphics[width= \textwidth]{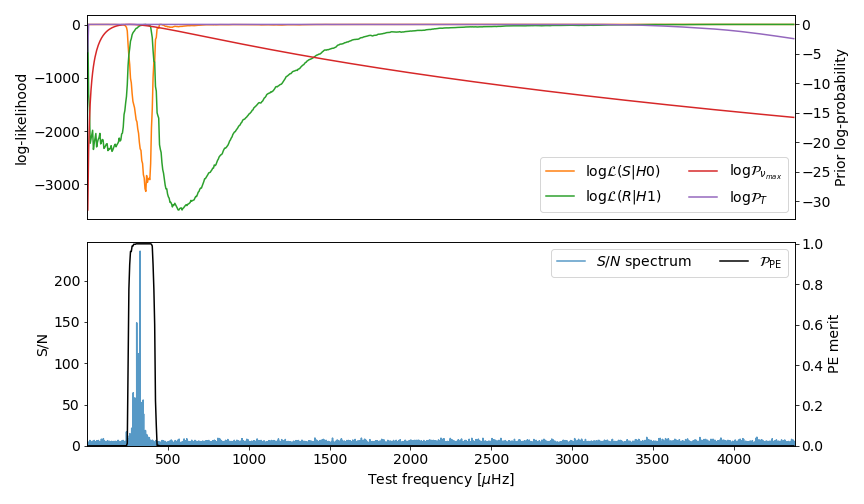}
    \caption{Example of the likelihoods and priors involved in computing the posterior probability $\mathcal{P}_{\mathrm{H}_1}$, for the star \epsret. The likelihoods and priors are shown on a log-scale for clarity, while the \snr and posterior are shown on a linear scale.}
    \label{fig:H0H1prob}
\end{figure*}

\subsubsection{Priors on $\mathrm{H}_0$ and $\mathrm{H}_1$} 
\label{sec:PEpriors}
We use two priors to compute $P_{\mathrm{PE}}$. The first estimates the probability that the predicted power exceeds a false alarm probability threshold. This requires that we compute a \snr threshold at each frequency that the predicted power must exceed for a false alarm probability of $1\%$ for example. At each test frequency we therefore solve
\begin{equation}
    \mathcal{P}_{\mathrm{thr},i} = \chi^2_{2\Nenvi}(S_{\mathrm{thr}}),  
\end{equation}
for $S_{\mathrm{thr},i}$, where $S_{\mathrm{thr},i}=\sum_{i-\Nenvi/2}^{i+\Nenvi/2} p_{\mathrm{thr},i}/b_i$ is the \snr threshold that the predicted power must exceed to yield a false alarm probability less than $\mathcal{P}_{\mathrm{thr}}=0.01$. Given the predicted power, $p_{\mathrm{pred}}$ we can then compute the probability that this threshold is exceeded by
\begin{equation}
    \mathcal{P}_{\mathrm{T},i} = \int_{R_{\mathrm{thr},i}}^{\infty}\chi^2_{2\Nenvi}(R)\, dR,
    \label{eq:pT}
\end{equation}
where $R_{\mathrm{thr},i} =\sum_{i-\Nenvi/2}^{i+\Nenvi/2} S_{\mathrm{thr},i}/ S_{\mathrm{pred},i}$, and $S_{\mathrm{pred},i}=p_{\mathrm{pred},i}/b_i$.  The effect of $\mathcal{P}_\mathrm{T}$ is to penalize test frequencies where the predicted power is so small that, given the background, we do not expect to see power that with certainty can be attributed to a \pmode envelope. 

The second prior estimates $\numax$ of the target from measurements that are independent of the power spectrum. We use  Eq. \ref{eq:numax}, \ref{eq:dnunumax} and the scaling relation for $\dnu$ \citep{Kjeldsen1995}
\begin{equation}
    \frac{\dnu}{\Delta \nu_{\odot}} = \left(\frac{M}{\Msun}\right)^{0.5} \left(\frac{R}{\Rsun}\right)^{-1.5},
\end{equation}
to obtain an approximate guess $\tilde{\nu}_{max}$
\begin{equation}
    \frac{\tilde{\nu}_{max}}{\numaxsun} = \left(\frac{R}{\Rsun}\right)^{\frac{0.5}{0.5-a}}\left(\frac{\Teff}{\Teffsun}\right)^{\frac{-0.25}{0.5-a}}.
    \label{eq:aproxnumax}
\end{equation}
Here $R$ and $M$ are the stellar radius and mass respectively. Using the \kepler training set consisting of \MS stars and low-luminosity \RG stars we found $a=0.791$. However, we note that $a$ varies on the percent level between \MS and evolved \RG stars, and so is not necessarily applicable to stars that are different from our training set.

To each frequency bin we can then assign a weight
\begin{equation}
    \mathcal{P}_{\numax,i} = \exp\left(\frac{-\log(\nu_i / \tilde{\nu}_{max})^2}{2\sigma_{\numax}^2}\right).
    \label{eq:pN}
\end{equation}
Equation \ref{eq:aproxnumax} neglects the mass dependence on $\numax$ since this is typically unknown a priori. The radius on the other hand may be more readily estimated using broadband photometry and parallax measurements. Based on the scatter of the residuals when applying Eq. \ref{eq:aproxnumax} to the \kepler training set we set $\sigma_{\numax}=0.5$. 

We can then finally compute the prior probability on $\mathrm{H}_1$ by
\begin{equation}
    \mathcal{P}_{\mathrm{H}_1,i} = 0.5\,\mathcal{P}_{\numax,i} \, \mathcal{P}_{\mathrm{T},i},
    \label{eq:priorH1}
\end{equation}
and thereby
\begin{equation}
    \mathcal{P}_{\mathrm{H}_0,i} = 1-\mathcal{P}_{\mathrm{H}_1,i}.
\end{equation}
This prior penalizes frequency bins where we either do not expect $\numax$ to be, based on independent observables, or where the predicted power of the oscillations would be too low to produce a signal that exceeds the false alarm probability. Equation \ref{eq:priorH1} is scaled such that it tends to 0.5 when the joint probability $\mathcal{P}_{T}\,\mathcal{P}_{\numax}\approx1$, which ensures that the prior simply excludes frequencies where we do not expect the envelope to be located or be visible. Figure \ref{fig:H0H1prob} shows the likelihoods and priors in relation to the posterior probability and \snr spectrum of \epsret, where the posterior probability is used as a merit function to evaluate the detection. 

\subsection{The repeating pattern detection method}
\begin{figure}
    \centering
    \includegraphics[width=1\columnwidth]{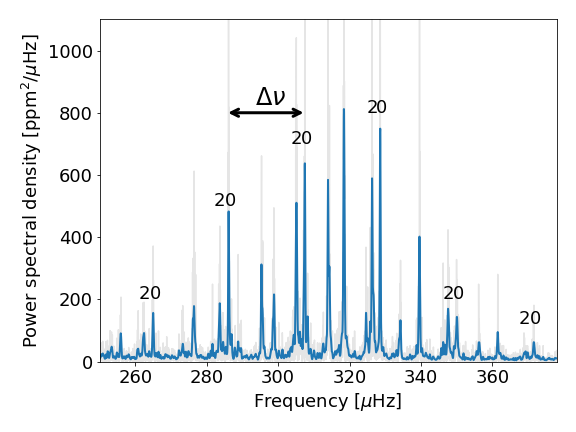}
    \caption{Example power spectrum of \epsret centered on the \pmode envelope. The smoothed and unsmoothed \PSD is shown in blue and gray, respectively. The labeled $l=2$ and $l=0$ modes appear in pairs over the range of the envelope, separated by the large frequency separation $\dnu$. The remaining $l=1$ and $l=3$ modes are not labeled for clarity.}
    \label{fig:dnu_example}
\end{figure}
In solar-like oscillators the large frequency separation, $\dnu$, between modes of the same angular degree but consecutive radial orders is approximately constant across the \pmode envelope (see Fig. \ref{fig:dnu_example}). The presence of such a regularly repeating pattern in a region of power excess is a strong indicator of solar-like oscillations. 

The large separation is sensitive to the mean density of the star and relates to the sound crossing time of a wave from one side of the star to the other, and back again. \citet{Roxburgh2006} suggested using the \ACF of the time series to exploit this feature of the \pmodes to measure the large frequency separation. \citet{Mosser2009b} extended this by applying a band-pass filter to the time series which spans the \pmode frequencies, allowing both $\dnu$ and $\numax$ to be precisely measured. 

Here we use the same method as a means of detecting the solar-like oscillations. As with the power excess method we test each frequency in the \PSD, so that when the band-pass filter overlaps the \pmode envelope, the \ACF shows a local maximum at an \ACF lag, $\tau$, equivalent to $1/\dnu$. In contrast, when the filter does not significantly overlap the envelope the resulting \ACF shows little to no response at $\tau=1/\dnu$ or at any other values of $\tau$. This feature may then be used to both detect the presence of a \pmode envelope, and localize it in the spectrum. 

For each test frequency we apply a band-pass filter $W$. As was done by \citet{Mosser2009b} we use a Hanning filter given by
\begin{equation}
    W_i=\begin{cases}
        0.5-0.5\cos\left(\frac{2\pi(\nu-\nu_i)}{\Nenvi}\right) &|\nu-\nu_i|\leq\gammaenvi/2\\ 
        0 &  |\nu-\nu_i|>\gammaenvi/2,
        \end{cases}
\label{eq:hanning}
\end{equation}
where the $\gammaenv$ is a function of the test frequency. 

The \ACF of the time series is then computed by the inverse Fourier transform of the filtered signal-to-noise spectrum by
\begin{equation}
    C_i(\tau) = \int_{0}^{\nu_{\mathrm{Nyquist}}}S(\nu)\,W_i(\nu) e^{j2\pi\nu\tau}d\nu, 
\label{eq:acf}
\end{equation}
where $j^2=-1$. Repeating this process for all the test frequencies that we want to investigate yields a two-dimensional complex array $G$ of the \ACF as a function of test frequency. The \ACF array shows a response at $\tau \approx 1/\dnu$ when the filter $W$ either partially or fully overlaps the \pmode envelope at $\numax$.

\begin{figure}
    \centering
    \includegraphics[width= 1\columnwidth]{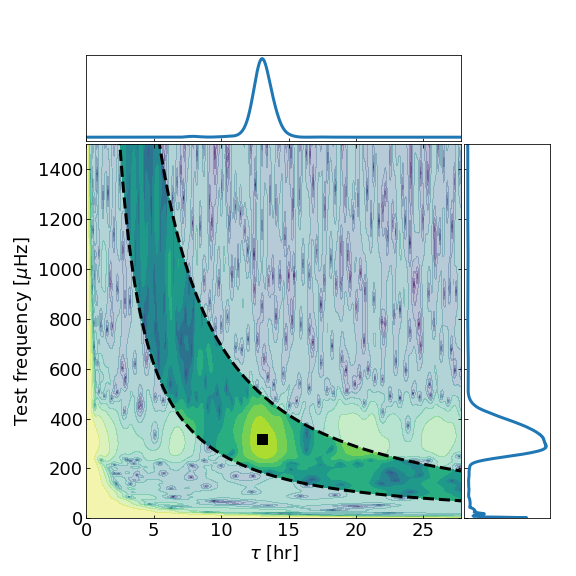}
    \caption{Contour map of the squared absolute value of the \ACF for \epsret. At each test frequency the \ACF is normalized to unity at $\tau=0$. The area between the dashed lines denotes the interval considered when estimating the detection metric. The solid black square shows the $\dnu$ and $\numax$ for this target. The top and right panels show the collapsed \ACF along both axes.}
    \label{fig:ACFmap}
\end{figure}

To establish a detection metric we compute the mean of the squared absolute value of $G$, normalized to unity at $\tau=0$ at each test frequency, so that 
\begin{equation}
    r_i = \frac{1}{N_{\tau,i}}\sum_{k=a_i}^{b_i}{\frac{|G_{i,k}|^2}{|G_{i,0}|^2}}.
    \label{eq:CACF}
\end{equation}
We only sum over a small number of bins, $N_{\tau,i}$, defined by $a_i=u\,\Delta_T/\widetilde{\dnu}_i$ and $b_i=u^{-1}\Delta_T/\widetilde{\dnu}_i$, where $\Delta_T$ is the observing cadence, and $\widetilde{\dnu}_i$ is the estimate of $\dnu$ at a given test frequency given by the scaling relation in Eq. \ref{eq:dnunumax}. We set $u=10^{0.2}$ so that the interval accounts for the variance in observed values of $\dnu$ in the training set around the scaling relation. The resulting interval is shown in Fig. \ref{fig:ACFmap} for \epsret. 

Using the collapsed \ACF we can establish a detection probability by computing the probability that the response due to an envelope is inconsistent with noise. By testing a sample of $10^4$ simulated white noise time series, we found that the resulting noise statistics of $r$ can be well approximated by a $\Gamma$ distribution. The detection probability can then be written as
\begin{equation}
     \mathcal{P}_{\mathrm{RP},i} = \left[\frac{\beta^{\alpha}}{\gamma{\left(\alpha\right)}} r_i^{\,\alpha-1} \exp\left(-\beta r_i\right)\right]^{-1},
\label{eq:RPprob}
\end{equation}
where the shape parameter $\alpha=\mu^2/\sigma^2$ and scale parameter $\beta=\mu/\sigma^2$. When applying the Hanning band-pass filter to a constant white noise time series, the mean of the collapsed \ACF at a given test frequency can be shown to be
\begin{equation}
\mu_i = \frac{3}{2}\frac{1}{\Nenvi\,N_{\tau,i}}.
\end{equation}
 
From the noise simulations we found that the variance can be approximated as
\begin{equation}
    \sigma_i^2 \approx \frac{\mu_i^2}{\Nenvi} \left(1+\frac{N_{T}}{N_{\tau,i}}\right),
\end{equation}
where $N_{T}$ is the length of the time series\footnote{For time series on the order of years, it may be prudent to bin the power spectrum for speed before computing the \ACF. In this case both $\mu$ and $\sigma^2$ should be scaled by the inverse binning factor.} in units of the observation cadence. Finally, we normalize $\mathcal{P}_{\mathrm{RP}}$ to range between $0$ and $1$, similar to the \PE posterior probability, such that we can establish a detection threshold.

We note that the probability from Eq. \ref{eq:RPprob} does not formally account for the correlation of $r$ between frequency bins due to the width of the Hanning filter. The resulting detection probability from Eq. \ref{eq:RPprob} can therefore only be considered approximate when applied to real \tess time series. However, from the testing set (see Sect. \ref{sec:performance}) we find that this effect is likely very small due to the low number of false positive detections obtained with the \RP method. 

\section{Performance}
\label{sec:performance}
\subsection{Optimal detection thresholds}

To evaluate the performance of the two methods we must first establish the criteria for a detection. The detection probabilities are in both cases normalized to range between 0 and 1. We therefore set the requirement that for a method to yield a detection the merit must  exceed a predetermined threshold at any test frequency in the spectrum. These thresholds are determined based on the testing set established in Sect. \ref{sec:tgtselection}. 

We evaluate the performance of the \PE and \RP methods in terms of the \TP and \FP rates, as well as the overall accuracy of the methods. A true positive is a detection in the oscillator test set, while an false positive is a detection in the nonoscillator set. The accuracy is the total number of true detections and true nondetections in the entire test set. These metrics are evaluated under two conditions: that a given set of thresholds yields a detection by either of the two methods, or by both. 

\begin{figure*}
    \centering
    \includegraphics[width=\textwidth, trim={3cm 1cm 3cm 0}, clip]{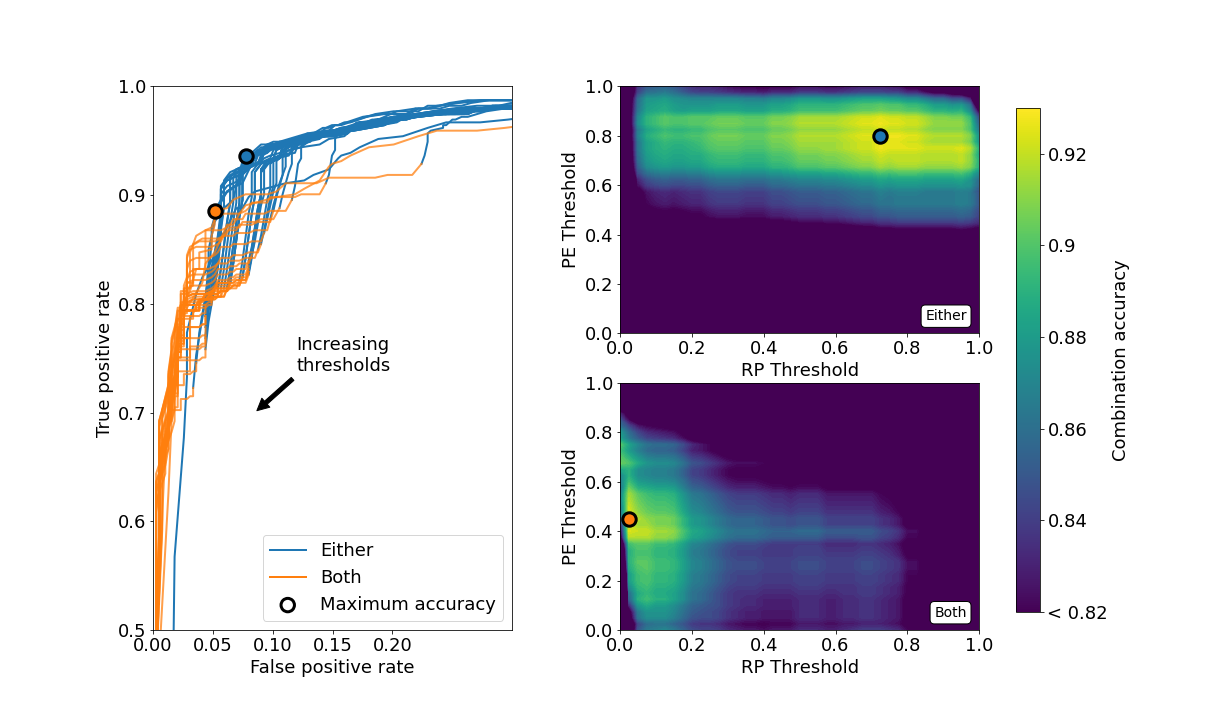}
    \caption{Performance summary for the \PE and \RP methods. Left frame: The \TP rate and the \FP rate for the different combinations of the \PE and \RP thresholds. The detection thresholds for both methods generally increase going from the high \TP and \FP range to toward low \TP and \FP. The two sets of curves denote \TP and \FP values when the requirement is that either (blue) or both (orange) the \PE or \RP method yields a threshold crossing. For $\mathrm{TP}<0.5$ the curves tend toward $\mathrm{FP}=0$, and similarly for $\mathrm{FP}>0.2$ they tend toward $\mathrm{TP}=1$. Right frames: the accuracy of the \PE and \RP methods under the two conditions. The circled dots show the values for the thresholds that maximize the accuracy of the methods under the two conditions. }
    \label{fig:AUC}
\end{figure*}

Figure \ref{fig:AUC} shows the recorded \TP and \FP rates and the associated accuracy for thresholds for the \PE and \RP methods ranging between $0$ and $1$. We vary the threshold for each merit function independently, yielding a set of possible receiver operating characteristic (ROC) curves for each of the conditions where either of the thresholds is exceeded, or both. In general, a combination of higher thresholds leads to a lower \TP rate, but also a lower \FP rate.

The maximum achievable accuracy of $93.2\%$ is found when setting a threshold of $0.77$ and $0.73$ for the \PE and \RP methods, respectively, when one requires only a detection in either of the methods. This yields a \TP rate of $94.7\%$ but with an \FP rate of $8.2\%$. When one sets the stricter requirement that both methods yield a detection, the highest accuracy of $91.5\%$ is found for a \PE and \RP threshold combination of $0.4$ and $0.03$ respectively. While this decreases the \TP rate to $88.8\%$, it reduces the \FP rate to $5.7\%$.  

\subsection{Characteristics of false positives and negatives}
\begin{figure}
    \centering
    \includegraphics[width=1\columnwidth]{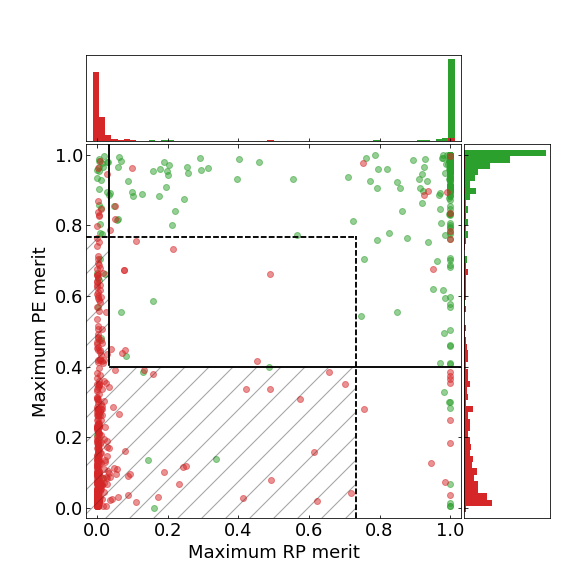}
    \caption{Maximum recorded values of the \PE and \RP figures of merit for the \tess oscillating test set (green) and the nonoscillating test set (red) described in Sect. \ref{sec:tgtselection}. Above the dashed line are targets with detections from at least one of the detection methods, and targets above the thick solid line yield a detection with both methods. Targets that fall in the hatched region are nondetections under both conditions.}
    \label{fig:tessmerit}
\end{figure}

The thresholds that yield the maximum accuracy can be used to evaluate which targets might tend to be incorrectly flagged as positive detections. Figure \ref{fig:tessmerit} shows the maximum recorded values of the two merit functions for the two samples in the testing set, as well as the regions corresponding to the two sets of maximum accuracy thresholds. Based on these thresholds we visually inspect  the false positives and false negatives from the two test sets to identify pathological cases. 

Of the false positive detections in the nonoscillating sample, we note that seven targets show a single, narrow, high \snr peak in the power spectrum. These are likely due to persistent systematic instrumental variability that is not correctly removed in the time series reduction, or another source of variability in the target or a background star. These peaks can produce a response in the \RP merit if the variability is long-lived, or in the \PE merit as well for very high \snr variability where the $\mathrm{H}_0$ likelihood becomes exceptionally small compared to $\mathrm{H}_1$.  

A further 23 false positive detections only show a marginal power excess in the \PSD, if any at all. Of these targets some only show a response in the \PE merit, which could be caused by variability that by chance produces a power excess which is similar to that of a \pmode envelope.  However, the majority of these targets also show an \RP response, and the prior estimates of $\numax$ are at least partially consistent. This leaves the possibility that these are oscillating stars that we cannot visually confirm, and so were labeled as nonoscillators during the manual validation.

We also identified two false positive cases which show multiple broad maxima in the \RP merit at several frequencies. Both of these targets show large changes in the time series variance between two separate observing campaigns, and so the assumptions about the statistics used to compute the \RP merits likely no longer hold, producing incorrect labeling. 

In the oscillating test sample we find that the gap removal reduces the observed \snr and thus the response on the \PE merit, but also in the \RP merit, leading to a reduced \TP rate. This could be caused by, for example, long lived modes that are made to appear incoherent after the gap removal, leading to a lower response in the \RP merit. However, despite the \snr reduction the gap removal dramatically reduces the \FP rate (see Fig. \ref{fig:keplermerit}) since the assumptions on the noise statistics for both methods would otherwise be much weaker. 

A similar reduction of the \snr may happen if the dilution of the stellar flux in the photometric aperture is inaccurate or unaccounted for. The effect of flux dilution is to wash out the variability of a star, which decreases the \snr below the predicted value. For our testing sample $90\%$ of the targets have dilution, $D>0.95$, with less than one percent variation between sectors, and is therefore likely not sufficient to affect our test sample. However, when observing more crowded fields closer to the galactic plane, for example, this might produce larger differences in the dilution between sectors. Revising the pixel masks to achieve a more consistent dilution estimate over time may therefore become necessary to avoid misclassifying potential oscillators.  Additionally, the presence of binary companions with a small angular separation from the oscillating star that fall within the same aperture may also contribute to a flux dilution.

Additional false negatives can occur when the prior on $\numax$ is significantly different from an observed envelope. In such cases the \PE merit may be reduced to below the detection threshold. We identify five such cases, where only one shows a nearby source bright enough to potentially contaminate the photometric aperture mask. This leaves the alternative that the effective temperatures, parallaxes and $K_{S}$ magnitudes, used to compute the $\numax$ prior might be inaccurate in these cases. Unresolved binary companions may contribute to the uncertainty in, for example, broadband photometry, thereby potentially skewing the resulting estimate of $\numax$. 

Lastly, we note that some targets may show a lower than expected power excess due to a lack of visible modes. This can occur if the period of the oscillation mode is particularly long compared to the length of the observations, and the mode is then considered unresolved. While a single sector of \tess observations is in principle sufficient to resolve all the oscillations expected in our test sample, the short time series also confer a significantly reduced \snr. For faint targets, single modes in the envelope may therefore not appear at all, producing a low observed power excess. Alternatively, targets which show depressed dipole mode amplitudes \citep{Mosser2012b, Stello2016} will also show a power excess which is less than the scaling relations predict. These can occur in evolved solar-like oscillators with masses $M\gtrsim1.1\Msun$, but their presence is difficult to predict based on a priori observed quantities. We therefore do not currently include any measures to account for depressed dipole modes. 
 
\section{Conclusions}
\label{sec:conclusions}
We have constructed an algorithm for detecting solar-like oscillations in light curves of \MS and \SG stars observed by \tess. The algorithm consists of two separate modules. The first computes a running average of the \PSD that scales with the width of an expected \pmode envelope, and assigns a detection probability to each test frequency depending on how well it compares with the asteroseismic scaling relations for the envelope power. The second method computes an \ACF of the band-pass filtered time series, where a statistically significant response that is consistent with $\dnu$ indicates the presence of a \pmode envelope. The two methods can be used individually or in combination. The results of the application of this detection algorithm to a larger sample of \tess data will be presented separately in Hatt et al. (2022, in prep). 

To test the performance of the algorithm we manually validated a set of 400 oscillating and 400 nonoscillating \MS and \SG stars observed by \tess. When the detection methods are applied under the condition that either of the methods yields a detection we achieve a maximum accuracy of $93.2\%$, with associated \TP and \FP rates of $94.7\%$ and $8.2\%$ respectively. Applying the stricter condition that both methods yield a detection, the algorithm achieves a lower accuracy of $91.5\%$, with a \TP rate of $88.8\%$ and \FP rate of $5.7\%$. This translates into obtaining $\approx30\%$ fewer false positive detections in a given sample of stars, at the cost of finding $\approx6\%$ fewer of the oscillators in the sample. 

An even lower \FP rate can be achieved by increasing the thresholds in the merit functions that are used to confirm a detection, but this depends on the purpose of the analysis. For example, given a small sample of stars that can be manually vetted, lower thresholds may be set to maximize the number of detected oscillators, at the cost of a high \FP rate. Conversely, for a large sample size, where even a moderate detection rate ensures a high yield, strict thresholds can be set to reduce the \FP rate which may otherwise be detrimental to any subsequently inferred sample statistics.

While \tess has yielded millions of light curves already, the observing strategy is not the most favorable for detecting solar-like oscillators. We find that the gaps left in the time series between observing sectors produces numerous false positive detections. We mitigate this by adjusting the time stamps of the flux measurements in the time series such that gaps longer than $50$ days are removed. For the upcoming \plato mission this will be less of a problem as the minimum pointing duration is likely to be $\approx90$ days rather than the $\approx27$ days for \tess observations. The observing strategy may also include single-field stare campaign of a year or more, only interspersed with short data download periods.  

Additionally, we also saw false positives due to systematic effects, either inherent in the observations or from the data reduction method. The \SPOC pipeline is optimized to fulfill the science objective of \tess, which is to detect exoplanets. The reduction methods are therefore not necessarily optimal for asteroseismic studies. Dedicated data reduction methods for asteroseismology like those employed by the \tess Data for Asteroseismology \citep[T'DA,][]{Lund2021}, will improve the yield of detections by removing contaminating effects. 

As they are presented here, the detection methods will need to be developed further to automatically flag binary systems. For binary systems with a mass ratio substantially less than one, where only a single component shows visible oscillations, the nonoscillating star might impact the detection by flux dilution in the aperture or by skewing the prior estimate on $\numax$ if it is based on, for example, broadband photometry. While rarer, binary configurations can occur where the mass ratio is close to unity, and both are massive enough to produce visible oscillations \citep[][]{Miglio2014}. In cases where the \pmode envelopes significantly overlap \citep{White2017a}, the power excess detection method would yield a lower detection probability since the envelopes will produce twice the expected power in the same frequency range. The repeating pattern detection module on the other hand will produce a clearer detection since both stars adhere to the scaling relation between $\dnu$ and $\numax$. If the envelopes are sufficiently separated \citep[see, e.g.,][]{Appourchaux2015}, both the power excess and repeating pattern methods will produce detections at two distinct frequency ranges. However, we currently only require a single frequency bin to cross the detection thresholds, which will not indicate multiple maxima in the detection probabilities and thus if a light curve contains a binary pair. More elaborate selection criteria are required to flag these as potential binaries.

We make use of data from the \glsunset{TIC}\TIC as a prior to inform the detection process. We found a small number of cases where the data used to construct the prior are potentially inaccurate which highlights an additional area for improvement. Inaccurate prior information might be mitigated by, for example, including multiple sources of broadband photometry to estimate the stellar brightness, to compute a joint probability density, and to estimate the stellar temperature and radius and, in turn, $\numax$. In addition, on a more basic level the scaling relations used throughout the algorithm may be replaced by nonparametric probability densities based on previous observations \citep[as in, e.g.,][]{Nielsen2021a}. This will more accurately reflect the change in the mean and variance of the envelope parameters between different types of stars, and can be easily updated as new observations become available. Constructing such a relation however requires a dense sample of stars across at several evolutionary stages, but the current and future observations provided by \tess will likely contribute significantly to this endeavor.

\begin{acknowledgements}
MBN, WHB, and WJC acknowledge support from the UK Space Agency. 

GRD, and WJC acknowledge the support of the UK Science and Technology Facilities Council (STFC). 

This paper has received funding from the European Research Council (ERC) under the European Union’s Horizon 2020 research and innovation programme (CartographY GA. 804752).

The authors acknowledge use of the Blue-BEAR HPC service at the University of Birmingham. 

This paper includes data collected by the Kepler mission and obtained from the MAST data archive at the Space Telescope Science Institute (STScI). Funding for the Kepler mission is provided by the NASA Science Mission Directorate. STScI is operated by the Association of Universities for Research in Astronomy, Inc., under NASA contract NAS 5–26555. 

This work has made use of data from the European Space Agency (ESA) mission {\it Gaia} (\url{https://www.cosmos.esa.int/gaia}), processed by the {\it Gaia} Data Processing and Analysis Consortium (DPAC, \url{https://www.cosmos.esa.int/web/gaia/dpac/consortium}). Funding for the DPAC has been provided by national institutions, in particular the institutions participating in the {\it Gaia} Multilateral Agreement.

This publication makes use of data products from the Two Micron All Sky Survey, which is a joint project of the University of Massachusetts and the Infrared Processing and Analysis Center/California Institute of Technology, funded by the National Aeronautics and Space Administration and the National Science Foundation.

This paper includes data collected by the TESS mission. Funding for the TESS mission is provided by the NASA's Science Mission Directorate.

\end{acknowledgements}

\bibliographystyle{aa}
\bibliography{main}

\begin{appendix}
\section{Additional performance measurements}
\FloatBarrier

\begin{figure}[!h]
\label{fig:gapeffects}
    \centering
    \includegraphics[width=1\columnwidth]{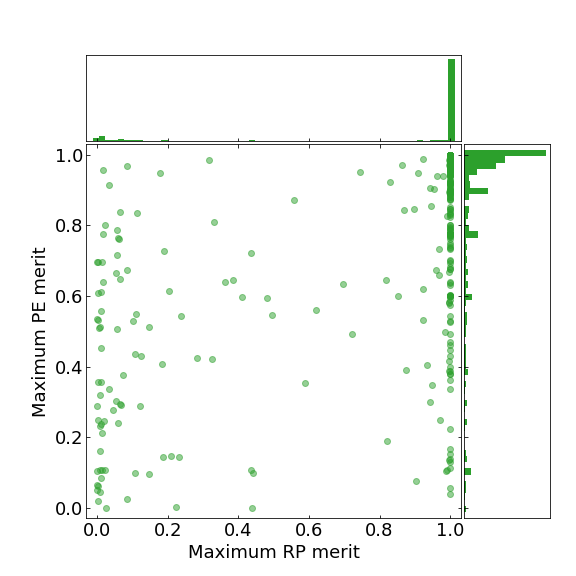}
    \includegraphics[width=1\columnwidth]{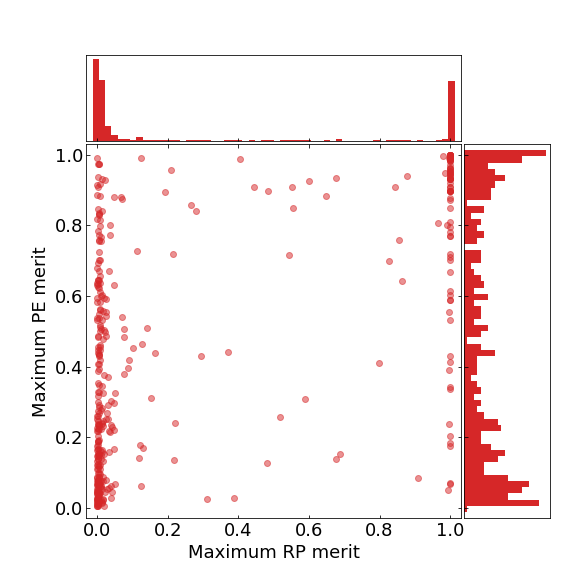}
    \caption{The maximum recorded values of the \PE and \RP figures of merit. Left: The recorded values for the \kepler short cadence sample with no gap removal. Right: The same as the left frame but for the \tess nonoscillating test sample, also without any gap removal.}
    \label{fig:keplermerit}
\end{figure}

\section{Deriving the predicted envelope power}
\label{app:ppred}
To compute the predicted power of a notional \pmode envelope for a target of a given $\Teff$, we must derive an estimate of the envelope height that is centered on any test frequency in the \snr spectrum. This derivation follows in part that shown in \citet[][chapter 5]{Basu2017}.

The notional envelope height in the \PSD can be expressed as
\begin{equation}
    \frac{\Henv}{H_{\mathrm{E},\odot}} = \frac{\Delta\nu_{\odot}}{\dnu}\left(\frac{\Amax}{A_{\mathrm{max},\odot}}\right)^2,
    \label{eq:henv}
\end{equation}
where $A_{\mathrm{max}}$ is the amplitude of a radial mode at $\numax$, and is given by \citep{Kjeldsen1995} 
\begin{equation}
    \frac{\Amax}{A_{\mathrm{max},\odot}} = \beta \left(\frac{L/\Lsun}{M/\Msun}\right)\left(\frac{\Teff}{\Teffsun}\right)^{-2}.
    \label{eq:amax}
\end{equation}
The stellar luminosity $L/\Lsun$, mass $M/\Msun$, and effective surface temperature $\Teff/\Teffsun$ are expressed in terms of the equivalent solar quantities, where we use the same solar values at in \citet{Schofield2019}. In the \tess band-pass $H_{\mathrm{E},\odot}\approx0.1 \ppmmuHz$ and $\Amax \approx 3\,\mathrm{ppm}$ .

We implement a correction factor, $\beta$, for the observed reduction of the mode amplitudes for stars that lie near the red edge of the $\delta$ Scuti instability strip. This correction is given by
\begin{equation}
    \beta = 1-\exp{\left(-\frac{T_{\mathrm{red}}-\Teff}{\Delta T}\right)},
    \label{eq:beta}
\end{equation}
where 
\begin{equation}
    T_{\mathrm{red}} = T_{\mathrm{red},\odot}\left(\frac{L}{\Lsun}\right)^{-0.093},
    \label{eq:tred}
\end{equation}
which is the temperature at the red edge of the $\delta$ Scuti instability strip for a given luminosity, where $T_{\mathrm{red},\odot}=8907$K \citep{Houdek1999}. We use $\Delta T=1550$K as was determined by \citet{Chaplin2011a}. 

 When using Eq. \ref{eq:beta} and \ref{eq:tred}, $\beta$ may become $\approx 0$ for a notional envelope placed at a low test frequency. The predicted \snr will therefore appear equivalent to the background, leading to a falsely high detection probability. However, this typically only occurs around a frequency of approximately $10\muHz$, which is much lower than the range that we are concerned with for \MS and \SG stars. However, if this method is applied to \RG stars with a lower $\numax$, this should be taken into account.  

In order to compute the predicted $\Amax$ we rewrite \ref{eq:amax} using $L\propto R^2\Teff^4$ and the scaling relation for $\numax$ \citep[e.g.,][]{Kjeldsen1995}
\begin{equation}
    \frac{\numax}{\numaxsun} = \frac{M}{\Msun}\left(\frac{R}{\Rsun}\right)^{-2}\left(\frac{\Teff}{\Teffsun}\right)^{-0.5},
\label{eq:numax}
\end{equation}
such that
\begin{equation}
    \Amax = A_{\mathrm{max},\odot}\, \beta\, V\,\left(\frac{\numax}{\nu_{\mathrm{max},\odot}}\right)^{-1} \left(\frac{\Teff}{\Teffsun}\right)^{1.5}.
\end{equation}
Here we introduce the correction factor $V$, which adjusts the mode amplitude to that expected in the \tess photometric bandpass. For \kepler observations $V=1$, and for \tess we use $V=0.85$ as computed by \citet{Campante2019}, based on the method by \citet{Ballot2011b}.

Observations of $A_{max}$ show a degree of variance around Eq. \ref{eq:amax} \citep[see, e.g,][]{Huber2011, Lund2017}. We assume that this variance is approximately Gaussian around $\log{\Amax}$ at a given value of $\numax$, and so can be accounted for by
\begin{equation}
    \tilde{A}_{max} = \frac{1}{\sqrt{2\pi\sigma_A^2}}\exp{\frac{-\log({A^{\prime}/A_{max}})^2}{2\sigma_A^2}},
\end{equation}
where $A^{\prime}$ is a parameter that allows us to marginalize over the amplitude variance around the scaling relation. We use $\sigma_A=0.1\,\mathrm{ppm}$ based on the results in \citet{Huber2011}.

To rewrite $\Henv$ in terms of $\numax$ alone, we can use the approximation \citep[see, e.g.,][]{Mosser2012} 
\begin{equation}
\frac{\Delta \nu}{\Delta \nu_{\odot}} = \frac{\numax}{\nu_{\mathrm{max},\odot}}^{a}. 
\label{eq:dnunumax}
\end{equation}
The exponent $a=0.791$ is determined based on a linear fit to the parameters of the training set defined in Sect. \ref{sec:tgtselection}.
We can then write the envelope height as
\begin{equation}
    \Henv = \,\eta^2 D^2 H_{\mathrm{E},\odot} \left(\frac{\numax}{\nu_{\mathrm{max},\odot}}\right)^{-\alpha}  \tilde{A}_{max}^2.    
\end{equation}
We have applied two additional correction terms: $\eta^2$, and $D^2$. The first correction, $\eta^2$, reduces the predicted envelope height due to the attenuation of an oscillation observed at discrete time intervals, and is given by
\begin{equation}
    \eta = \mathrm{sinc}\left(\frac{\pi\numax}{2\nu_{Nyquist}}\right).
\end{equation}
The second correction, $D$, is due to the dilution of the flux in the aperture of the target star caused by nearby sources such as bright background stars or binary companions. For an isolated star with only faint background stars, $D\approx1$, but this may be less in crowded fields for example. We use the crowding factor as described by \citet{Thompson2016} for \kepler data, and \citet{Twicken2020} for \tess as representative of the flux dilution $D$. For targets observed in multiple sectors, the crowding factor may change due to reorientation of the pixel mask. In such cases we average the crowding factor to estimate the dilution. 
        
We can now write the predicted power as
\begin{equation}
    p_{\mathrm{pred}} = \Henv \exp\left(-\frac{(\nu - \numax)^2}{2c^2}\right),
    \label{eq:ppred}
\end{equation}
where
\begin{equation}
    c=\frac{\gammaenv}{2\sqrt{2\ln{2}}}.
\end{equation}
  
\end{appendix}
\end{document}